# Health and cleanliness of the XMM-Newton science payload since launch

M.G.F. Kirsch[1a], A. Abbey[c], B.Altieri[a], D. Baskill[c], K. Dennerl[b], J.van Dooren[e], J. Fauste[a], M.J. Freyberg[b], C. Gabriel[a], F. Haberl[b], H. Hartmann[d], G. Hartner[b], N. Meidinger[b], L. Metcalfe[a], B. Olabarri[a], A.M.T. Pollock[a], A.M. Read[c], S. Rives[a], S. Sembay[c], M.J.S. Smith[a], M. Stuhlinger[a], A. Talavera[a]

[a]European Space Agency (ESA), European Space Astronomy Center (ESAC), Villafranca, Apartado 50727, 28080 Madrid, Spain, [b]Max-Planck-Institut für extraterrestrische Physik, Giessenbachstrasse 1, 85748 Garching, Germany, [c]Dept. of Physics and Astronomy, Leicester University, Leicester LE1 7RH, U.K., [d]EADS Astrium GmbH, [e]ESA, ESTEC, The Netherlands

**ABSTRACT**

On December 10th 2004 the XMM-Newton observatory celebrated its 5th year in orbit. Since the beginning of the mission steady health and contamination monitoring has been performed by the XMM-Newton SOC and the instrument teams. Main targets of the monitoring, using scientific data for all instruments on board, are the behaviour of the Charge Transfer Efficiency, the gain, the effective area and the bad, hot and noisy pixels. The monitoring is performed by combination of calibration observations using internal radioactive calibration sources with observations of astronomical targets. In addition a set of housekeeping parameters is continuously monitored reflecting the health situation of the instruments from an engineering point of view. We show trend behaviour over the 5 years especially in combination with events like solar flares and other events affecting the performance of the instruments.

## 1. INTRODUCTION

Scientific instruments operated in the unfriendly environment outside the atmosphere of the Earth suffer from extreme conditions due to radiation, temperature and low pressure. All three factors and their combinations can cause rapid aging and foster contamination of the instruments. This can lead, in the case of radiation for example, to effects ranging from a degradation in Charge Transfer Efficiency (CTE) of the CCDs, up to a fatal failure of the whole instrument. Should contamination occur, the effective area of the instruments will be affected lowering the efficiency of response and introducing additional absorption edges in the measured spectra. Tracking the health and cleanliness status of the system at all stages of the mission is clearly fundamental to ensuring sustained performance of the instruments and to preserving the quality of the scientific data. Vigilant monitoring is linked to appropriate safety and recovery procedures in critical cases, and in general can be expected to lead to instrument calibration refinements and refined scientific data output. XMM-Newton[1] was launched in December 1999 on an Ariane 504 rocket from French Guyana. Its six instruments are operated in parallel through the course of a 48-hour highly elliptical orbit. Three Wolter type-1 telescopes with 58 nested mirror shells focus X-ray photons onto the five X-ray instruments comprising the European Photon Imaging[2,3] Camera (EPIC) and the Reflecting Grating Spectrometers[4] (RGS). The Optical Monitor[5] (OM), employing a 30 cm Ritchey Chrétien optical telescope, can perform parallel optical observations of the same field. EPIC is comprised of three cameras employing two distinct detector technologies. The two EPIC-MOS cameras use front illuminated EPIC-MOS (Metal-Oxide Semi-conductor) CCDs as X-ray detectors, while the EPIC-pn camera is equipped with an EPIC-pn (p-n-junction) CCD which has been specially developed for XMM-Newton. EPIC provides spatially resolved spectroscopy over a field-of-view of 30' with moderate energy resolution. The EPIC cameras can be operated in various observational modes related to the specific readout strategies in each mode. For a detailed description of the modes see Kendziorra et al. (1997)[6], Kendziorra et al. (1999)[7], Kuster et al. (1999)[8] and Ehle et al. (2003)[9]. The RGS is designed for high-resolution spectroscopy of bright sources in the energy range from 0.3 to 2.1 keV. The OM extends the spectral coverage of XMM-Newton into the UV and optical, and thus opens the possibility to test physical models of source spectra against data over a broad energy band. The six OM filters allow colour discrimination, and there are two grisms, one operating in the UV and one in the optical, to provide low-resolution spectroscopy.

---

[1] mkirsch@sciops.esa.es +34 91 8131 345; fax +34 91 8131 172

## 2. WHAT DOES HEALTHY AND CLEAN MEAN?

### 2.1. Instrument internal – Radiation

CCD's are sensitive detectors of radiation. That holds true whether the radiation is electromagnetic, from the infrared to the gamma ray range, or ionising particle radiation. The CCDs of XMM-Newton's X-ray instruments are designed to respond to X-ray radiation in the energy range from 0.1-15 keV. Nevertheless, they also detect particles, and photons in other energy ranges. We therefore divide the radiation detected by the XMM-Newton CCDs into 4 categories: X-ray radiation, optical radiation, "soft"-particle radiation, and "hard"-particle radiation. Optical radiation seen by the detectors has to be taken into account in the scientific analysis of XMM-Newton data, but neither does it change the properties of the devices with time. The critical component, therefore, from the point of view of instrument aging, is the particle radiation. The soft component consists of low energy protons ("soft" protons), with energies ranging up to 300 keV, that are funnelled by the XMM-Newton mirrors under grazing incidence to fall on the detectors in just the same way as X-ray photon events. Upon hitting the CCDs the protons can damage the silicon lattice and the electrode structures at the surface of the CCDs. This is problematic for the front-illuminated EPIC-MOS CCDs, leading to reduction of the Charge Transfer Efficiency (CTE) and degradation of the energy resolution. To protect the EPIC-MOS cameras from those soft protons the filter wheel is moved to a closed position in periods of high soft-proton flux. The term "hard"-particle radiation refers to cosmic background radiation particles and associated secondary particles which are not focussed by the mirrors, and may consist of protons, nuclei or other Minimally Ionising Particles (MIPs). These can also cause damage to the instruments. Regardless of their intrinsic incident energy MIPs deposit $\sim 2$ MeVg$^{-1}$ in target material. A MIP can easily penetrate the camera boxes from any direction and deposit energy in the CCD.

### 2.2. Contamination

A contaminant is any material in the whole light path of an instrument that should not be there and which affects the efficiency of the instrument. This can be, in the simplest case, frozen water showing-up as oxygen features in spectra, or it might be any other substance that might out-gas from the spacecraft and freeze onto any of the instruments. The two main categories of contamination are particulate contamination, which has almost energy-independent impacts on the effective area, and molecular contamination, the effects of which are energy dependent.

#### 2.2.1. Molecular contamination

Molecular contaminants typically consist of hydrocarbons or water. The hydrocarbon contamination levels are expected to be of the order of $5\text{-}10\cdot 10^{-8}$ g.cm$^{-2}$ for smooth surfaces under atmospheric conditions in the cleanest environments (XMM-Newton-SOC-PS-TN-0003). With a typical density of 1 g cm$^{-3}$ these values correspond to a monolayer thickness of 5-10 Å. Simulations from the work reported in the above technical note show that for molecular contamination with hydrocarbons the reflectance of the hydrocarbon film over the gold coating is similar to that of a bilayer made up of a low and a high Z material. Indeed the reflectivity increases at low energies and is gradually less affected towards higher energies. For energies below the Au-M absorption edge at 2.2 keV the change in absolute reflectivity is about 0.2 % per Å of hydrocarbon.

| OPTICAL ELEMENT | GRAZING ANGLE | PFO (PPM) FOR 1 % EFFECTIVE AREA REDUCTION |
|---|---|---|
| Grating | 2.275 deg | 200 |
| Mirror 1 | 39.7 arcmin | 29 |
| Mirror 24 | 26.6 arcmin | 21 |
| Mirror 58 | 17.4 arcmin | 13 |

Table 1: Reduction of effective area due to Particle Fall Out (PFO) (Note that the unit ppm relates to surface coverage. 1ppm means that in total 1 mm² of 1 m² is covered or shadowed by particles)

#### 2.2.2. Particulate contamination

Particles absorb and diffractively scatter X-rays. Reflective scattering can be neglected at X-ray energies and with a good approximation it can be considered that grains absorb X-rays passing through their geometrical cross-section. The dust fractional coverage of a mirror surface is related to the Particle Fall Out (PFO). That typically depends on the exposure time of the mirror surface in a clean room. Table 1 gives estimates of the PFO that would lead to a 1 % change in effective area. Although the change of effective area due to particulate contamination is almost energy independent, slight energy dependence is introduced since effective area vignetting by dust depends on grazing angle. At high energies, where only the inner mirror shells contribute to the effective area, the impact of particulate contamination on relative effective area variation can be expected to be higher by almost a factor of 2 than at low energies, where the collecting area of the outer mirror shells dominates. Table 1 shows also clearly that mirrors are much more sensitive to contamination than gratings due to their shallower grazing angles.

### 2.2.3. Cleanliness Requirements

Pre-launch limits for contamination levels of key components in the optical path were:

| REQUIREMENTS | PARTICULATE CONTAMINATION | MOLECULAR CONTAMINATION |
|---|---|---|
| Mirror Modules | < 70 ppm | < $8 \cdot 10^{-8}$ g/cm$^2$ |
| Reflecting Grating Assembly | < 350 ppm | < $4 \cdot 10^{-8}$ g/cm$^2$ |

In order to fulfil the requirements of an absolute effective area accuracy of 10 % at any energy between 0.1-10 keV and a relative accuracy of better than 3 %, until the end of the XMM-Newton lifetime, a maximum of 200 ppm for particulate contamination and $2 \cdot 10^{-7}$ g.cm$^{-2}$ for molecular contamination should not be exceeded. Note that after an XMM-Newton contamination working group meeting in November 1995 the contamination budget for the mirror shells was revised anticipating that contamination levels could reach 140 ppm of particulate contamination and $1.5 \cdot 10^{-7}$ g cm$^{-2}$ of molecular contamination between the end of the ground calibration tests at the PANTER test facility and the start of the in-orbit operations. This is a factor of 2 above what was foreseen in the original requirements.

## 2.3. The importance of contamination monitoring

Contamination on instruments can significantly change the detection efficiency. This may for example lead to a change in effective area. Where the change is detected and can be quantified the calibration can be adjusted and scientific data quality is not greatly affected. Contamination, if it is substantial, may be revealed through its impacts on scientific results, but where it is slight it may be very hard to detect, since it may mimic a slightly higher astrophysical absorption column density in astronomical spectra. Therefore it is crucial to perform systematic contamination measurements before and after launch in order to fully track the contamination history.

## 2.4. Contamination prevention

In the design phase the cleanliness requirements have been accounted for by making the mirror modules, the telescope tube and the experiments separately closed units having their own doors and purging devices. The tube and the mirror modules had to be kept closed, except for relatively brief moments during optical testing. Special mirrors were used for alignment such that the flight mirror units did not have to be opened. All mirror modules, the telescope tube and the optical monitor have been continuously purged with pure nitrogen or synthetic air. The EPIC cameras were evacuated while the RGS cameras were pressurised with nitrogen. The telescope tube is sealed from Carbon Fibre Reinforced Plastic (CFRP) out gassing towards the inside by a continuous aluminium foil, acting as a vapour barrier. In addition the aluminium foil is covered for stray-light suppression with a 25 μm thick black kapton foil that is glued with low out-gassing adhesive. After thermal-vacuum testing in the Large Space Simulator (LSS) at ESTEC, tape-lifts measured 50 ppm average inside the tube, respectively < $0.1 \cdot 10^{-7}$ g cm$^{-2}$ from wipe tests.

# 3. XMM-NEWTON OPERATIONS - 5 YEARS IN ORBIT

## 3.1. Thermal history

The EPIC-pn-CCD camera aboard XMM-Newton shows a correlation between the gain in the signal processing chain and the temperature of the quadrant boxes, which house the CCD electronics. A temperature decrease of 1 °C corresponds to an energy shift of ~0.4 adu (~ 2 eV) at the Mn-K$_\alpha$ energy of 5.9 keV (Dennerl 2001[16]). Figure 1 shows the EPIC-pn quadrant box temperature for quadrant Q1 (where the on-axis CCD is located) for all available Full-Frame mode data, averaged over an exposure. The step-like feature originates in the intrinsic resolution of the temperature sensor data acquisition. The line is a linear regression to most of the data (during revolutions 120-140 sometimes EPIC-MOS or RGS was switched off and the temperature environment was different for this known reason, the fit uses only revolutions later than 150). There is a clear trend, seen in fact for all modes, with a temperature decrease of about 3.2 mK per revolution (~ 0.59 °C per year).

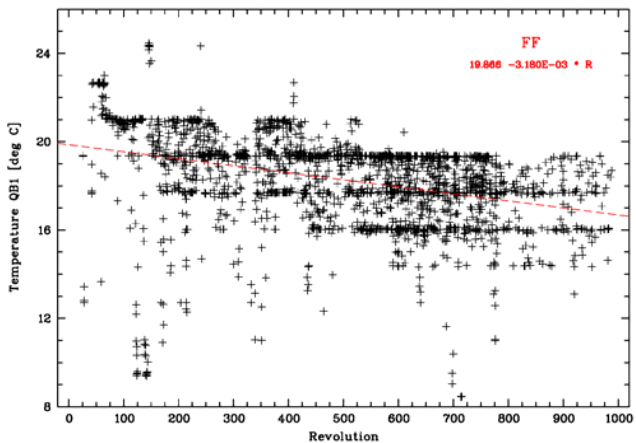

Figure 1: Evolution of the EPIC-pn quadrant box temperature

A similar trend has independently been found for the EPIC-MOS electronics boxes and also for the total satellite temperature. It has been suggested that the known orbit drift and consequently reduced reflected emission from Earth decreased the temperature. Up to now no significant drift in line energies of Al-K, Mn-$K_\alpha$, and Cu-K in EPIC-pn could be identified. However, it should be noted that a small ($< 10^{-5}$) systematic long-term variation of the EPIC-pn oscillator frequency has been found which could be related to the temperature variation described here (Freyberg et al. 2005[17]).

As another example of the housekeeping monitoring the temperature of all focal plane CCDs is shown in Figure 2. As a consequence of the cooling of the EPIC-MOS and RGS instruments (see Section 3.2) these temperatures drop at the end of 2002. Figure 2 shows, in addition, the excursions of the temperature due to eclipse seasons when the cameras are switched off during spacecraft passages through the Earth's shadow and are, for safety reasons, heated with the substitution heaters. This causes an increase of the temperature. Moreover a seasonal effect is seen that causes a kind of wave structure in temperature with a maximum around Christmas. The increase of the temperatures only happens at the end and the beginning of the revolution and is due to the influence of the albedo of the earth. It is assumed that scientific data is not affected by those temperature excursions since scientific observations are only started when the nominal CCD temperatures are reached.

### 3.2. Cooling

During November and December 2002 a series of operations were performed to lower the nominal operating temperatures of the RGS and EPIC-MOS instruments, and to subsequently re-establish their calibration. The purpose of the cooling was to move important detector characteristics back towards the values which they had earlier in the mission. Parameters such as CTE (and therefore energy determination and energy resolution), and number of hot pixels etc. had deteriorated since launch in the hostile space radiation environment.

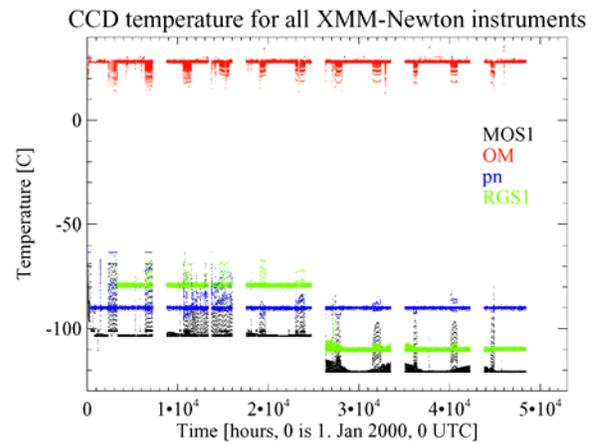

Figure 2: Temperature of the XMM-Newton focal plane CCDs for all instruments. The data gaps at the end of one year are due to system memory problems at the time of recreating the data.

#### 3.2.1. Cooling the EPIC MOS cameras

While ground-based software corrections for EPIC-MOS-CTE managed to recover most of the energy loss, inevitably the detector line-profile FWHM had widened due to imperfect correction and statistical noise of charge trapping. Since the EPIC-MOS CCDs are qualified to operate down to -130 °C it was expected that an improvement would be seen in CTE at reduced temperatures (optimised at –120 °C). Many bright pixels had developed, as expected, due to radiation and micrometeoroid impacts. It was expected that, also for those pixels, cooling should have had a curative effect. The two EPIC-MOS instruments were cooled in the night of November 6-7 2002, with the instruments in Closed/CalClosed configuration. The instruments were cooled from the previous operating temperature of -100 C, down to –120 C. The EPIC-MOS1 and EPIC-MOS2 cameras were cooled in parallel. The new temperature value was reached within 2.75 hours. The Charge Transfer Inefficiency (CTI=1-CTE) changed by a factor of 2-3 depending on the CCD considered. The serial CTI was unchanged. Prior to the cooling the CTE had tended to degrade, with much of the degradation occurring discontinuously in association with radiation bursts from solar flares. Following the cooling the CTE degradation evolves with the underlying slope, a trend unbroken even after strong solar flares in 2003 and 2004. That means that the detectors are much more robust against radiation at a temperature of -120 centigrade. The energy resolution was improved by 13% and 17% for EPIC-MOS1 and EPIC-MOS2 respectively at Mn energies (6 keV) and by about 10 % at Al energies (1.5 keV), varying somewhat among the CCDs. The number of bad pixels in the Current Calibration Files was reduced by a factor of 2.6 (EPIC-MOS1) and 7.0 (EPIC-MOS2). The few remaining ones are mostly the defects from micrometeoroid impacts. We saw no change in the quantum efficiency following the cooling. Observations on the isolated neutron star RXJ 0720.4-3125 and the Vela SNR before and after cooling substantiate that statement. We performed, after cooling, a contamination monitoring using the SNR N132D, the isolated neutron star RXJ 0720.4-3125 and the Vela supernova remnant. There was no sign of post-cooling adsorption of contaminants onto the detectors, which had been the main perceived risk factor in the process. With this new thermal set point (-120 °C), the EPIC-MOS focal plane is now sensitive to the thermal impact of the Earth at every perigee passage. Consequently, the focal plane temperature can increase to –115 °C at perigee. Note that the amplitude of the thermal excursion also depends on the relative alignment of Sun/Earth/Satellite. From an operational point of view it is important that the EPIC-MOS focal plane temperature is stabilized at –120 °C at the start of the observation window. Further studies have shown that this requirement is met even in the worst case. Nevertheless, in case a further cooling is to be performed in the future this operational constraint will have to be taken into account.

### 3.2.2. Cooling the RGS spectrometers

The RGS CCDs were cooled to a new operating temperature of -110 °C. The main result of cooling for RGS was a huge reduction in the incidence of flickering pixels. The CTI was restored to close to its pre-flight values. Previously poor CTI near the edges of the detectors, where it had been 100 times worse than for the inner regions, was almost eliminated. The CCD noise was greatly reduced, and the width of the noise distribution peak became, and remains, narrower than before launch. There is now just one hot column and one hot pixel in RGS2.

### 3.3. Radiation history, seasonal radiation

The high fluctuations and variability of the particle background in the cameras was one of the main surprises and concerns after launch. It can be divided into two main components: a) A rather quiescent component, due to the interaction of high-energy particles (energies larger than about 100 MeV) with the structure surrounding the detectors and possibly with the detectors themselves, and its associated fluorescence, and b) A variable and flaring component attributed to soft-protons (with energies smaller than a few 100 keV), which are presumably funnelled toward the detectors by the X-ray mirrors and hence form a vignetted background. This flaring component of the soft-protons was not foreseen before launch. The spectra of soft-proton flares are variable.

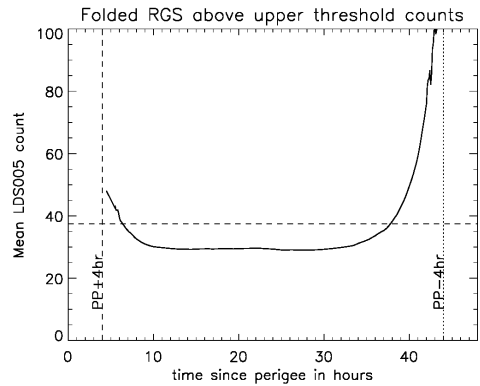

Figure 3: Average level of soft-proton radiation events seen by RGS through the XMM-Newton revolution is plotted against time since perigee passage (PP). This parameter is a good indicator of soft-proton flaring. It shows that the probability of high background periods is higher toward the perigee, with higher radiation at the beginning and end of the revolution, but with significant asymmetry: Higher radiation is found on the descending part of the orbit. On average, good observing time is found between PP+6 hours and PP-10 hours. The dashed horizontal line is 5 sigma above the quiescent level.

The probability of encountering soft-proton flaring decreases with altitude (distance from Earth), as for higher-energy radiation detected by the Radiation Monitor (a few MeV) (see Fig. 3). Hence this population of soft-protons must be linked, to some extent, to the radiation belts, but in a much more unpredictable and loose way than is the case for higher-energy particles. A higher fraction of the descending part of the orbit (in-bound leg, toward perigee) is affected by radiation, as can be seen in Fig. 3, due to the offset and tilt of the geomagnetic axis with respect to the Earth's rotational axis.

A seasonal effect can also be seen due to the one-year rotation of the magnetosphere and magneto-tail in the XMM-Newton orbit reference frame (revolution of two sidereal days): the portion of the orbit crossing the magneto-tail and/or the night side of the magnetosphere is more prone to radiation flaring. In spring the descending part of the orbit is more affected, while in autumn the ascending part is affected. These two geometric effects lead to asymmetries in the radiation profile along the XMM-Newton orbit. Overall, up to 15-20 % of the XMM-Newton-observing time (the 40 hours available for scientific observations above the radiation belts, out of the 48 hour orbit) can be lost, although this is difficult to assess quantitatively as the impact of radiation on science output depends on the scientific goal of an observation. This fraction of time lost is highly variable from revolution to revolution and observation to observation, but there is a higher probability that observations planned toward the end of the revolutions will be affected, where EPIC exposures have often to be aborted before their scheduled end times. The effect of soft-proton flaring is a bigger nuisance for XMM-Newton than for Chandra, for several reasons: Chandra's orbit brings it further from the Earth than XMM-Newton, so that it spends much more time above the region of quasi-trapped soft-proton clouds. XMM-Newton's effective area (i.e. capacity to collect X-rays) is much higher, by a factor of 5-10 depending on energy. It therefore not only collects more X-ray photons, but also focuses more soft-protons into the focal plane.

### 3.4. EPIC-MOS patch redistribution

A small patch of anomalous responsivity has been found on the central CCD of each MOS camera by examining all archived observational data for the source1ES0102. In addition, a raster scan has been performed to identify positional and temporal variability of the patches. The patches have degraded over time. The locations of the patches are coincident with the nominal position of sources when placed at the two effective pointing boresights for XMM-Newton: the EPIC-pn and RGS boresights. The patterns coincide with the peak in received photon dose of the detectors. The patches cause a significant change in the low energy redistribution characteristics of the EPIC-MOS cameras, which is spatially and temporarily dependent. However the situation seems to have stabilised and the characteristics of the patches are not consistent with being caused by a contamination mechanism. Epoch dependent canned Response Matrices (RMFs), currently under test, will convey the characteristics of the patches to the user community for application in their data analysis. The task *rmfgen*, for release with v6.5 of the XMM-Newton Scientific Analysis Software (SAS), will be able to produce spatially and epoch dependent RMFs for all epochs and off axis angles.

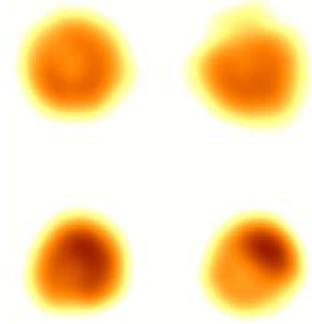

Figure 4: Smoothed images of the SNR 1ES0102 in the 0.1-0.35 keV energy band. Left: EPIC-MOS1 Right: EPIC-MOS 2. Upper rev 247. lower rev 447

### 3.5. Impacts on CCDs

EPIC cameras have suffered, so far, 4 events interpreted as micrometeoroid impacts in five years of in-orbit operation. These occurred in revolutions 107 (EPIC-MOS2), 156 (EPIC-pn), 325 (EPIC-MOS1), and 961 (EPIC-MOS1), and are understood to be the result of dust particles coming down the boresight direction and interacting with the mirror shells, so that they or the associated debris are channelled by the mirror shells onto the focal plane. The most recent event (discussed in more detail in Section 3.5.1 below) was by far the worst, leading to the loss of EPIC-MOS1 CCD6 (one of the 7 CCDs in the EPIC-MOS1 focal plane) in March 2005. For EPIC-MOS the impacts were characterized by an optical flash and the occurrence of FIFO data buffers full, and the sudden appearance of several high energy and high-recurrence frequency hot pixels. In the EPIC-pn case several new hot pixels also appeared within 1 frame time (Large Window mode, 48 ms) again pointing toward the impact of a dust micro-particle. RGS has not been affected so far, possibly because of the diffraction geometry leading to an off axis position of the RGS focal plane with respect to the mirror shells and the diffraction gratings. Assuming that the acceptance solid angle is about the same for dust particles as for X-rays, with reflections occurring for angles of incidence up to 3.5 degrees off-axis, one can derive a telescope effective aperture of 21.5 $cm^2$Sr for an effective area of 1500 $cm^2$, and hence one can estimate the rate at which particles may hit the focal plane using the micrometeoroid mass spectrum measured by Grün et al. (1985)[10] and Divine (1993)[11]. The typical particle mass matching the observed rate of events is m~$10^{-13}$ g, corresponding to a particle size of 0.2-0.3 μm. These particles are either interplanetary or interstellar dust, but are not linked to meteor showers, which do not contain particles smaller than about $10^{-8}$ g.

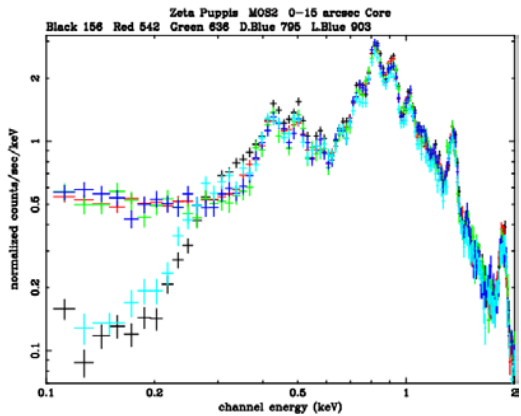

Figure 5: EPIC-MOS zeta pupis observations: at the patch location on rev. 156 (black), at the patch location on rev. 542 (red), patch location rev. 636 (dark blue), off-patch location rev. 903 (light blue)

### 3.5.1. Latest and severest EPIC-MOS impact

As introduced above, at about 01:30 hrs. UT on 09 March, 2005, during XMM-Newton revolution 961, an event was registered in the focal plane of the EPIC-MOS1 instrument. The characteristics of the event were reminiscent of very similar, but less energetic, events registered in the EPIC-MOS1 focal plane on September 17, 2001, the EPIC-MOS2 focal plane on August 12, 2002, and the EPIC-pn focal plane on October 19, 2000, which have been attributed to micrometeoroid impacts scattering debris into the focal plane. The consequences of the recent event have been the most significant. In the period immediately following the light flash it became apparent that EPIC-MOS1 CCD6 was no longer recording events, and that all CCD6 pixels were, in effect, returning signal at the saturation level. Standard recovery procedures were immediately applied, but had no effect on the observed behaviour, leading to the eventual conclusion that CCD6 had sustained terminal damage. New hot pixels were also recorded elsewhere in the EPIC-MOS1 focal plane.

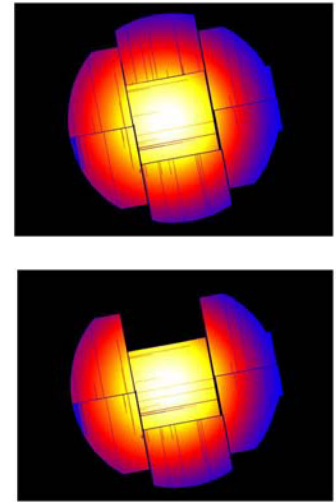

Figure 6:
Exposure maps corresponding to exposures before (upper) and after (lower) the loss of MOS1 CCD6

### 3.5.2. EPIC-pn impact

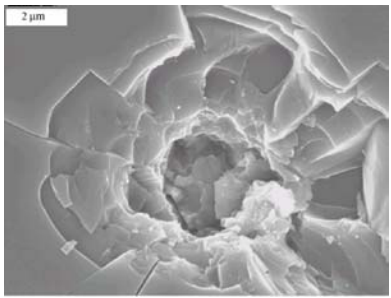

Figure 7: Experimental verification of micrometeoroid impact as result of the experiment at the dust accelerator. The SEM image shows the top view of a crater in silicon caused by a particle scattered off the mirror surface under grazing incidence

During revolution 156 about 35 individual bright pixels lit up out of approximately 150000 pixels of the 6 cm x 6 cm EPIC-pn detector area. The amount of leakage current generated in the pixels cannot be explained by single heavy ion impacts. The only reasonable possibility found to explain the observation was that a micrometeoroid[13], scattering off the mirror surface under grazing incidence, reached the focal plane detectors and produced the bright pixels. This proposal was studied experimentally, on the ground, at the Heidelberg dust accelerator. Micron-sized iron particles were accelerated to speeds of the order of 5 km s$^{-1}$ impinging on the surface of an X-ray mirror under grazing incidence[14].

The experiments verified that a dust particle under these conditions can cause damage consistent with the behaviour of the focal plane CCD-detectors of XMM-Newton. All the characteristics of the event and the damage could be reproduced in the laboratory. The formation of the bright pixels was caused by fragments of the primary dust particle.

## 4. INSTRUMENT PERFORMANCE MONITORING

### 4.1. Monitoring flow

Since XMM-Newton is in continuous contact with its Science Operations Centre (SOC) at the European Space Astronomy Centre (ESAC), near Madrid in Spain, we can perform a real time health monitoring. The health monitoring of the instruments is carried out in two steps. A so called Instrument Controller (INSCON) monitors the instrument performance 24 hours a day. Suspicious behaviour of the instruments is immediately reported to the instrument teams at ESAC and the PI institutes for further investigation. Further monitoring is then pursued offline by instrument-dedicated teams.

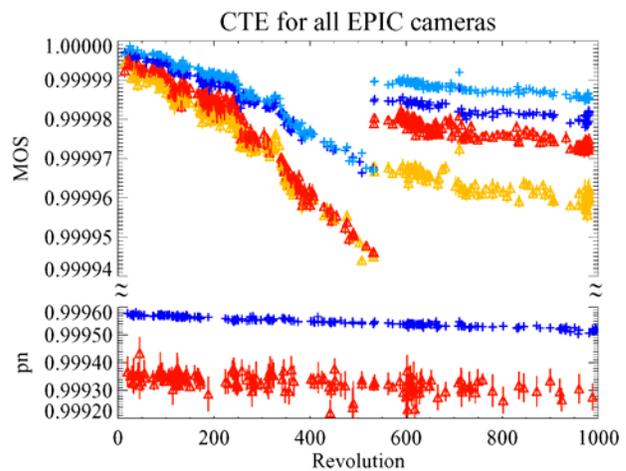

Figure 8:
Evolution of the CTE of the EPIC camera CCDs for the energies of the internal calibration source at Mn, 5896 eV, (blue crosses) and Al, 1486 eV (red triangles). Upper panel: EPIC-MOS, Lower panel EPIC-pn. The different blue and red tones in the upper pannel of the figure represent the EPIC-MOS1 (light blue (Mn), orange (Al)) and EPIC-MOS2 (dark blue (Mn), red (Al)). The discontinuity around rev.533 is related to the cooling of the EPIC-MOS cameras. Note that all EPIC-MOS CCDs have recovered some CTE after the cooling. Only the EPIC-MOS1 boresight CCD has not recovered as strongly - as shown by the orange crosses.

## 4.2. EPIC

Since launch the EPIC instrument consortium and the XMM-Newton SOC have performed an off line monitoring of the EPIC camera. It is separate from and complementary to the real time monitoring performed at the SOC by the INSCON team. The monitoring tools employed are largely automatic supporting the trend analysis of parameters which affect instrument performance and health. The main monitoring inputs are the regularly performed exposures with the filter in CalClosed position. Parameters monitored are charge transfer efficiency (CTE), energy resolution and gain, through the monitoring of the measured energies and widths of the internal calibration source lines. These parameters are essential for the reconstruction of photon event energy and system response, and are important diagnostics of instrumental health and performance. They therefore require close monitoring in order to detect, anticipate and possibly correct for larger than acceptable variations.

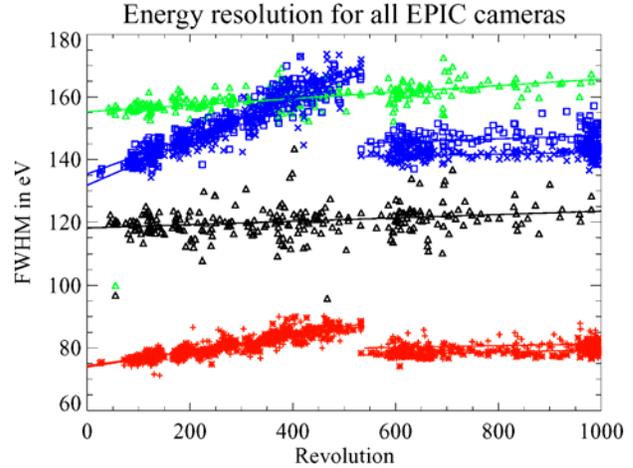

Figure 9:
Energy resolution of the central CCDs of the EPIC camera for different energies. EPIC-pn Al: black triangles, EPIC-pn Mn-K green triangles, EPIC-MOS1(2) Al: red crosses (asterisks), EPIC-MOS1(2) Mn-K: blue squares(x). Note that the improvement in energy resolution at revolution 533 is related to the cooling of the EPIC-MOS cameras (see 3.2.1). Note that the absolute values for Al for the EPIC-pn camera are slightly higher (5-10 eV) than the one by Meidinger[15] since no BG modeling has been performed

It is known that harsh radiation conditions may induce the formation of electron traps in the detectors, thus degrading the CTE. According to well-defined criteria the filter wheel is put in a closed position during periods of high background radiation to protect the detector. Consequently, the degradation of the EPIC-pn CTE is slight and in agreement with pre-launch predictions. As opposed to the EPIC-MOS, there is no clear correlation between the EPIC-pn CTE degradation and proton flares. Figure 8 shows the evolution of the CTE for the different EPIC cameras. Solar flares created a series of jumps in the CTE of the EPIC-MOS cameras prior to the cooling described above, while the EPIC-pn CTE degrades independently of solar flares at a nearly constant rate per year as seen in Fig. 8. Figure 9 shows the evolution of the energy resolution. It is clearly seen that the EPIC-MOS energy resolution at low energies is much better than that of EPIC-pn. At high energies, and prior to the cooling of the EPIC-MOS cameras, their energy resolution was degrading towards the level for the EPIC-pn. After cooling they have again a better energy resolution, also for high energies. As already described in Section 3.2, there is a correlation between EPIC-pn gain and the temperature of the electronics, and though satellite operations have been adapted to mitigate the temperature variations, this remains an important reason to monitor gain stability.

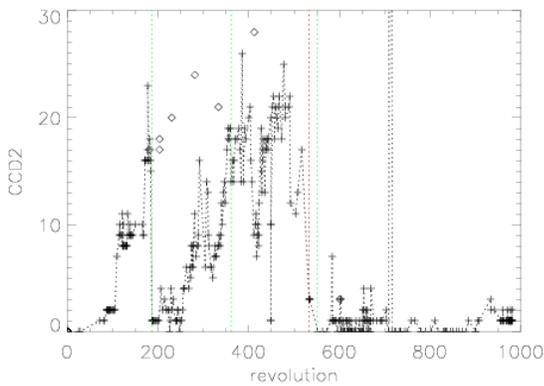

Figure 10:
Hot pixel evolution of EPIC-MOS2 CCD2. The diamonds show the absolute number of hot pixels. The crosses show the number of hot pixels that are not yet masked out on board (candidate hot pixels). The green lines indicate an update of the onboard bad pixel table that masks out hot pixels. Note that after the cooling in rev 533 most of the hot pixels have disappeared and the onboard bad pixel table could be relaxed to a few pixels per CCD.

By the term "bad" or "hot" pixel we mean any pixel within a CCD exhibiting abnormal behaviour which makes it useless for scientific data collection due to its tendency to mimic a signal (hot) or to yield no signal (bad). The number and location of bright pixels has to be monitored in order to flag pixels which have to be masked to reduce loading of the spacecraft telemetry budget, or because they adversely affect science quality. For the EPIC-MOS cameras the number of hot pixels increased through the mission due to micrometeoroid events (see Section 3.5) and due to aging caused by hard radiation particles. Figure 10 shows the evolution of hot pixels for CCD2 in EPIC-MOS2. For the EPIC-pn camera a small number of hot pixels, i.e. pixels with high dark current generation, were present at launch. Damage resulting from a suspected micrometeoroid impact in revolution 156 caused the sudden appearance of 35 additional hot pixels (see Section 3.5 above).

## 4.3. RGS

The RGS comprises two almost identical dispersive instruments, making it possible to follow different monitoring procedures to those required by the EPIC cameras. Each RGS has 9 CCDs, and one CCD from each instrument failed early in the mission due to problems with the associated electronics. However, the redundant design has ensured complete wavelength coverage between the two units. Two fixed radioactive monochromatic X-ray calibration sources illuminate parts of the detectors not illuminated by photons from the sky. Their signals are continuously recorded. Furthermore, scientific data are interleaved with diagnostic data to provide continuous simultaneous assessment of CCD characteristics. These data are separated from the scientific stream at an early stage of the data processing and subjected to independent analysis for monitoring CTI and event size and for the identification of the bad pixels and columns that are a routine feature of CCDs, in order to mask them out on-board and reduce telemetry load. These monitoring data are produced automatically and are available for inspection on the XMM-Newton website. In the early part of the mission, the RGS was operated at relatively high temperature in order to minimise the possibility of contamination due to out-gassing. This had the unwelcome consequence of increasing the incidence of noisy pixels. As mentioned above (Section 3.2) lowering the operating temperatures in November 2002 to nearer the design values resulted in a dramatic reduction of noise that has hardly increased in the 2½ years since, except to some extent in RGS1 CCD1. It is worth emphasising that the RGS is operated in parallel with the other instruments in every observation, which has made a large body of data available to characterise the variable particle background, for example, in apparently empty fields. The fixed calibration sources have decayed a few percent more rapidly than expected from their half-lives due probably to radioactive creep.

## 4.4. OM

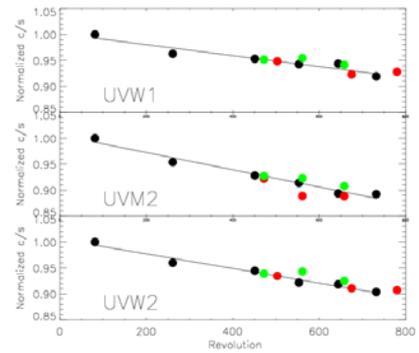

Figure 11: OM time sensitivity degradation in UV filters. Count rates for several standard stars at different epochs.

The operational conditions of the OM instrument (temperatures, currents, housekeeping parameters in general) are checked continuously by the operations control system. Some specific operations are carried out at activation of the instrument in each revolution. In particular, the so-called "eng-6" mode gives a diagnostic of the correct overall response of the OM. Some parameters are derived from this operational mode by the Instrument Controller (INSCON) and checked against required limits. Other specific operations performed more or less regularly are the measurement of dark current and flat field of the detector. The results of these are monitored off-line, after a quality control performed by the INSCON. During normal observations, housekeeping data are monitored continuously. In addition, all science files received from the instrument are inspected by the INSCON to search for possible anomalies. The background in the images is measured manually and recorded in a database for stability control. The performance (calibration) of the OM is monitored regularly. For this purpose a set of standard stars, whose spectral energy distributions are well known, is observed periodically. The count rates produced by the stars are plotted against time. In this way we have observed since launch time dependent sensitivity degradation, which is also wavelength dependent. This may be due to degradation of the detector, in particular of its S-20 photo-cathode. This degradation can be seen in Fig. 11 for the UV filters of OM (212 nm, 231 nm, 291 nm). A degradation of up to 2.8 % per year is observed in these filters, while in the optical ones (344 nm, 434 nm, 543 nm) it is less than 2 % per year.

## 5. CONTAMINATION MONITORING

### 5.1. The need for X-ray standards

There is just as much need in the X-ray domain, as at longer wavelengths like UV and optical, for a set of objects to serve as photometric or spectrometric standards. At optical and UV wavelengths, observations of absolutely calibrated celestial sources (normal stars, white dwarfs) set the standard. Their absolute calibration derives from a combination of (a) direct transfer of calibration by measurement from a few fundamental references, and ultimately from the reference Vega calibration and laboratory standards and (b) physical models of various stellar types, linked to the reference calibrations. X-ray equivalents are much harder to define, not least because variability is a very common property of X-ray sources. In most of the history of X-ray astronomy the Crab nebula has served as a reference which, in common with other fainter SNRs, is confidently expected to be constant, though the contribution of the Crab pulsar, while generally small, may complicate the matter in some high-energy regimes. With technological developments, increases in sensitivity and spatial and spectral resolution have generally gone hand-in-hand with decreasing field-of-view and detectors that do not operate optimally (that begin to saturate) for the brightest sources, such as the Crab. With these constraints, XMM-Newton has repeatedly observed a small set of objects that have been judged to serve as X-ray standards with the intention of both monitoring the evolution of instrument performance and providing straightforward comparisons with other instruments. Some of these objects have proved better than others. Kirsch et al.[12], give a modern view of the Crab. More promisingly however, although some isolated neutron stars have unexpectedly proved to be variable, several observations of RXJ1856-3754 with EPIC-pn have shown no discernable changes as discussed below. This object looks to be the best candidate X-ray standard identified so far, although its weakness and softness compromise its use for many instruments, including the RGS which has concentrated instead on bright SNRs in the Magellanic clouds whose distance has the advantage of limiting their angular sizes.

### 5.2. RXJ1856-3754

The isolated neutron star RXJ1856-3754 is used as a target to monitor contamination on the EPIC cameras. It has a very soft spectrum and is therefore well suited to measure possible contamination which would be expected to affect the low energy regime most strongly. Figure 12 shows three EPIC-pn SW mode spectra (all thin filter), which were fitted simultaneously with an absorbed blackbody model (constant*wabs*bbody) with common $N_H$, kT and normalization to energies between 0.15 and 1.2 keV. Only a constant factor relative to the first spectrum was allowed. The resulting parameters are kT: 62.3 ± 0.2 eV and $N_H$: 7.1 ± 0.3 ·$10^{19}$ cm$^{-2}$. The factors are: 1.0, 1.008 ± 0.005, 0.996 ± 0.005, i.e. fluxes deviate by less than 1 % or less than 1.6 sigma. These observations can be used to derive upper limits for any additional contamination between Revolutions 427 und 968 for carbon and oxygen as given in Table 2. In addition to that the SNRs N132D and 1ES0102 are used to measure contamination and stability of the energy calibration of the EPIC cameras. This analysis showed (see also Section 3.4) that the EPIC-MOS cameras have changed in their redistribution characteristics, but not in a manner consistent with contamination.

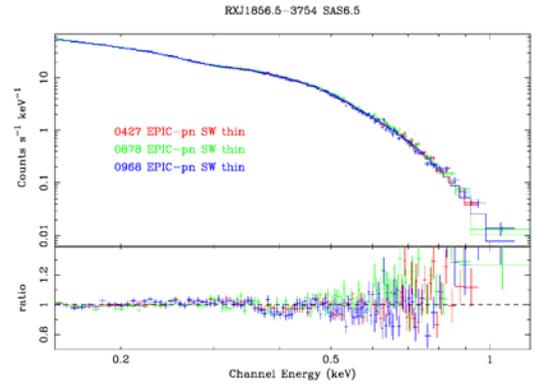

Figure 12: RXJ1856-3754 is used as a contamination monitoring target for the EPIC instruments. No contamination can be detected so far.

| Camera | Carbon | Oxygen |
|---|---|---|
| EPIC-pn | < 2.7e-07 g.cm$^{-2}$ | < 2.5e-06 g.cm$^{-2}$ |
| EPIC-MOS | < 7.2e-07 g.cm$^{-2}$ | < 1.3e-05 g.cm$^{-2}$ |

Table 2: Contamination upper limits for the XMM-Newton optical path as of rev. 427

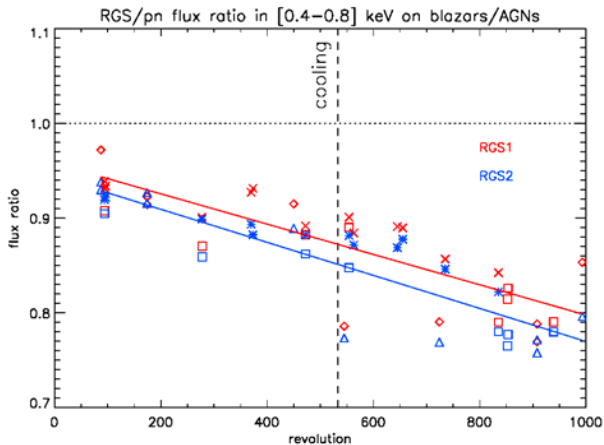

Figure 13:
Flux ratio of the RGS1 (red) and RGS2 (blue) versus EPIC-pn in the 0.4-0.8 keV range, for various sources with a continuous spectrum.

Since measurements indicate that the EPIC-pn camera is the most stable instrument on XMM-Newton and not contaminated we can use also variable objects to compare the relative fluxes with reference to EPIC-pn in order to measure time variability. PKS2155-304 and 3C273 have been used to perform a long-term comparison of the EPIC-pn camera with the RGS1 and RGS2 cameras as shown in Fig. 13. We see here a clear trend of decreasing flux in both RGSs by ~ 20 % (see also 5.3).

## 5.3. Monitoring the stability of RGS with SNRs and other objects

The high spectral resolution of the RGS instruments allows attention to individual spectral features for monitoring purposes throughout its bandwidth of 6-38Å, in most of which both RGS1 and RGS2 have proved stable, although recently evidence has started to emerge of changes in the longest wavelength sensitivity, such as that shown above in Fig. 13. That this is occurring well above the oxygen edge at 23.5 Å is confirmed by Fig. 14, which shows the ratio of continuum spectra of Mkn421 early and late in the mission. For a smooth spectrum, taking the ratio allows removal of the constant oxygen absorption features that are due to a combination of interstellar gas and instrumental material used in the manufacturing process: any changes due to a build-up of oxygen-rich contaminants would show up in this plot as a further edge - for which there is little or no evidence (see Fig. 14). The wavelength region between 6 and 25Å is rich in emission lines of many abundant elements and repeated measurements of various cosmic sources continue to confirm the lack of variability in most of the range. At longer wavelengths, emission lines are sparser and it becomes more difficult to select suitable objects. The spectrum of the bright O-star ζ Puppis has strong NVI lines near 29Å but the star has proved to be variable at about the 10% level. SNR 1ES0102-7219 in the SMC is of ideal size and confidently expected to be constant but shows a remarkable absence of lines between the OVII triplet near 22 Å and a weak CVI Lyman α emission line at 33.7 Å. Despite this weakness, its CVI line has been used as a long-wavelength stability monitor and the data are under active investigation. The particular analysis shown in Fig. 15 reveals an apparent decline in the line flux in both RGS1 and RGS2 of about 30 % over the course of the mission. In contrast the OVII and OVIII lines were constant within 1%. The amplitude of the CVI variation is dependent on the particular data selections employed which may indicate a connection with evolution of CCD charge-transfer properties rather than contamination. In any event, oxygen contamination is excluded though it is conceivable that carbon could be responsible, though the carbon edge falls outside the RGS waveband.

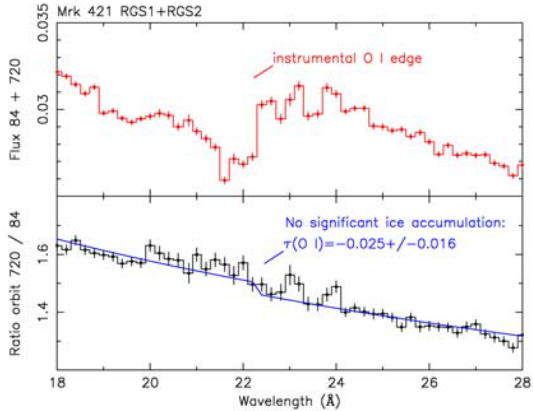

Figure 14: Ratio of continuum spectra of Mkn421 early and late in the mission

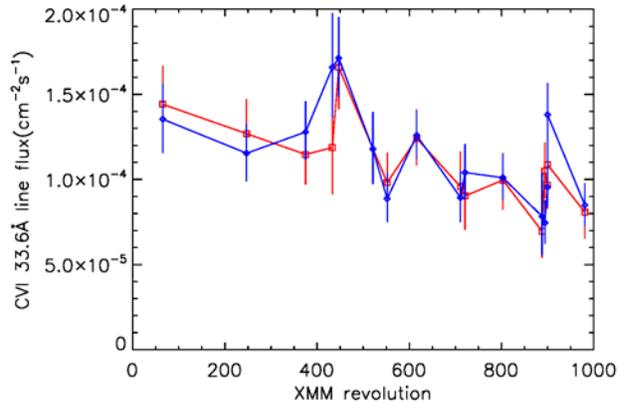

Figure 15:
CVI flux of the SNR 1ES0102-7219 in RGS1 (red) and RGS2 (blue)

## 5.4. OM

Laboratory measurements of all Optical Monitor components (reflectance of mirrors, transmission of filters, quantum efficiency of detector chain,....) allowed us to predict the throughput of the OM system. When XMM-Newton was launched, the in-flight throughput measured by observing standard stars (see Section 4.4) was found to be lower than expected, in particular in the UV filters. The deficit observed in the in-flight throughput, as low as 16 % at 212 nm, is independent from the time sensitivity degradation of the OM detector described in Section 4.4, which is much smaller.

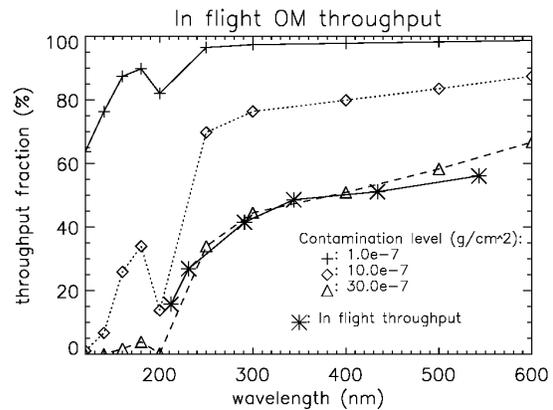

Figure 16: OM throughput with molecular contamination. In-flight throughput is represented with stars and the thick line.

Molecular contamination occurring prior to, or during, launch could be the reason for this deficit. If we consider the number of surfaces where the contaminating material can be deposited (9 for OM), then a density of $30 \cdot 10^{-7}$ g cm$^{-2}$ can account for the deficit in transmission. This is presented in Fig. 16, where the OM in-flight throughput derived from standard stars is compared with the predictions obtained by considering several densities of the contaminant material.

## 6. CONCLUSION

The EPIC cameras show no significant particulate or molecular contamination and only upper limits can be given, that are in line with the expectations from the contamination estimations on ground. The RGS spectrometers show a significant flux decrease near the carbon edge that, however, is probably not related to contamination because no effect is seen at oxygen energies. This decrease may instead be connected with evolution of CCD charge-transfer properties. The OM bandpass shows significantly reduced sensitivity relative to the expectation due probably to contamination that could have occurred prior to or during launch.

The higher soft proton flux first discovered by Chandra was clearly a surprise for both satellites and needs to be especially taken into account for future missions that will operate front illuminated CCDs. It is obvious that systematic internal calibration source measurements are mandatory for any such future mission. These should provide a complete sampling of the CTE behaviour over time in light of the observed impact of soft protons on CTE degradation of the front illuminated CCDs. On the other hand back-illuminated CCDs like those used in the EPIC-pn, proved to be robust against this effect.

The rate of micrometeoroid impacts suffered by XMM-Newton could be a significant issue for future missions with large collecting area.

In addition we see the strong need for a set of standard calibration sources for the X-ray regime. Given the luxury of having 6 satellites (XMM-Newton, Chandra, RXTE, Swift, Integral, Suzaku), that have X-ray instruments as their payload, in orbit simultaneously and for the coming years, we strongly recommend to form an international calibration group that may steer the cross-calibration efforts in this field.